\documentclass[aps,prd,onecolumn,showpacs,showkeys,amsmath,amssymb]{revtex4}
\usepackage{amsfonts}
\usepackage{amsmath}
\usepackage{graphicx}
\usepackage{subfigure}
\usepackage{dcolumn}
\usepackage{bm}
\usepackage{booktabs}
\usepackage[dvips]{color}
\makeatletter
\newcommand{\figcaption}{\def\@captype{figure}\caption}
\newcommand{\tabcaption}{\def\@captype{table}\caption}

\newcommand{\Rmnum}[1]{\expandafter\@slowromancap\romannumeral #1@}
\def\hlinewd#1{%
  \noalign{\ifnum0=`}\fi\hrule \@height #1 \futurelet
   \reserved@a\@xhline}
\makeatother
\def\qq{\langle\bar qq\rangle}
\def\qqb{\langle g_s\bar qq\rangle^2}
\def\GGa{\langle G^2\rangle}
\def\GGb{\langle g_s^2GG\rangle}
\def\GGG{\langle G^3\rangle}
\def\GGGb{\langle g_s^3fGGG\rangle}
\def\qGq{\langle\bar qg_s\sigma Gq\rangle}

\def\f(s){[(\alpha+\beta)m_c^2-\alpha\beta s]}
\def\non{\\ \nonumber}
\usepackage{ulem}

\begin{document}

\title{Masses of the tensor mesons with $J^{P}=2^{-}$}

\author{Wei Chen}
\email{stechen7@gmail.com}
\author{Zi-Xing Cai}
\affiliation{Department of Physics
and State Key Laboratory of Nuclear Physics and Technology\\
Peking University, Beijing 100871, China  }
\author{Shi-Lin Zhu}
\email{zhusl@pku.edu.cn} \affiliation{Department of Physics
and State Key Laboratory of Nuclear Physics and Technology\\
and Collaborative Innovation Center of Quantum Matter\\
and Center of High Energy Physics, Peking University, Beijing
100871, China }

\begin{abstract}

We calculate the two-point correlation function using the
interpolating current with $J^{PC}=2^{-}$. After performing the
Borel sum rule analysis, the extracted masses of the $2^{--}$
tensor charmonium and bottomonium are $3.97\pm0.25$ GeV and
$10.13\pm0.34$ GeV respectively. For comparison, we also perform
the moment sum rule analysis for the charmonium and bottomonium
systems. We extend the same analysis to study the
$\bar qq, \bar qs, \bar ss, \bar qc, \bar sc, \bar qb, \bar sb$
and $\bar cb$ systems. Their masses are $1.78\pm 0.12, 1.85\pm
0.14, 2.00\pm 0.16, 2.86\pm 0.14, 3.01\pm 0.21, 5.66\pm0.33,
6.40\pm 0.25$, and $7.08\pm 0.34$ GeV respectively.

\end{abstract}



\maketitle

%================================================================================
%================================================================================
\section{Introduction}\label{sec:Introduction}
%================================================================================
%================================================================================

Charmonium spectroscopy provides a crucial test of the quantum
chromodynamics(QCD) in both the perturbative and nonperturbative
regimes. The charmonium spectrum can be calculated using the
potential models ~\cite{2004-Brambilla-p-}. In the picture of the
conventional quark model, the charmonium states are characterized
by the $J^{PC}$ quantum numbers: $P=(-)^{L+1}, C=(-)^{L+S}$, where
$L$ is the orbital angular momentum and $S$ the total spin. Their
quantum numbers are $J^{PC}=0^{\pm+}, 1^{\pm-}, 1^{++}, 2^{++},
2^{-\pm}$ and so on.

There has been a revival of the charmonium spectroscopy in the
past eight years. Since the operation of the large facilities such
as Tevatron and two B-factories, many unexpected charmonium or
charmonium-like states above the open charm threshold have been
discovered, such as $X(3872), Y(4260), Y(4360), Y(4660),
Z^+(4430)$ \textit{etc.} ~\cite{2006-Swanson-p243-305,
2008-Zhu-p283-322, 2009-Bracko-p-, 2009-Yuan-p-,
2007-Rosner-p12002-12002}. Some of them tend to decay into a
charmonium state plus light hadrons. The underlying structures of
these new states are not known precisely. They are sometimes
speculated to be the candidates of the exotic states such as the
molecular states, the tetraquark states, the hybrid charmonium,
the baryonium states and so on.

After the discoveries of $\eta_c(2S)$ and
$h_c(1P)$~\cite{2008-Choi-p142001-142001,
2005-Rosner-p102003-102003}, all the charmonium states below the
open charm threshold have been observed experimentally
~\cite{2010-Nakamura-p75021-75021}. They are $J/\psi(1^3S_1),
\psi(2S)(2^3S_1), \chi_{c0,1,2}(1^3P_{0,1,2}), \eta_c(1^1S_0),
\eta_c(2S)(2^1S_0)$ and $h_c(1^1P_1)$. All these charmonium
resonances are narrow. Below the $D\bar D$ threshold, there is
only one tensor meson $\chi_{c2}(1P)$.

In the study of the meson spectroscopy, the local meson
interpolating fields $\bar{\psi}(x)\Gamma\psi(x)$ are usually
introduced. However, these operators are useful only in the study
of the low-lying states with $J^{PC}=0^{\pm+}, 1^{\pm-}, 1^{++}$
\textit{etc}. In order to explore the higher spin states, the non-local
fermion operators with covariant derivatives acting on the quark
fields should be used. For example, the authors studied the
charmonium spectrum including the higher spin states by including
the non-local operators in the framework of the lattice QCD
simulations in Ref.~\cite{2008-Dudek-p34501-34501}. The masses and
decay constants of the ground states heavy $\chi_{Q2}$ tensor
mesons have been calculated by using the QCD sum rules approach in
Ref.~\cite{2010-Aliev-p164-167}. The newly observed resonance
$X(1600)$ was studied as a $J^{PC}I^G=2^{++}2^+$ four-quark state
in the framework of QCD finite energy sum rules in
Ref.~\cite{2006-Wei-p4617-4626}. The tensor currents with
$J^{PC}=2^{++}$~\cite{1982-Reinders-p125-125, 1982-Aliev-p401-401}
and $2^{--}$ ~\cite{1982-Aliev-p401-401} were firstly employed to
study the light quarks systems.

In this work, we use the tensor current with $J^{PC}=2^{--}$ to
calculate the two-point correlation function in the most general
situation. Then we perform the QCD sum rule analysis and extract
the masses of the $2^{--}$ tensor states. Especially for the
charmonium and bottomonium states, we also perform the moment sum
rule analysis for comparison. None of these possible resonances
has been observed up to now except for the strange mesons
$K_2(1770)$ and $K_2(1820)$, which have $I(J^P)=\frac{1}{2}(2^-)$
with no definite $C$ parity.

The paper is organized as follows. In Sec.~\ref{sec:QSR}, we
discuss the quantum numbers of the interpolating current and
calculate the two-point correlation function in the general
situation. In order to cross-check the quantum number of the
interpolating current, we discuss the reduction of the current in
the non-relativistic limit in Appendix ~\ref{appendix1}. We
perform the Borel sum rule analysis for various systems in
Sec.~\ref{sec:numerical}. For comparison, we present the moment
sum rule analysis for the charmonium and bottomonium systems in
Sec.~\ref{sec:MSR}. The last section is a short summary.

%%%%%%%%%%%%%%%%%%%%%%%%%%%%%%%%%%%%%%%%%%%%%%%%%%%%%%%%%%%%%%%%%%%%%%%%%%%%%%
%================================================================================
%================================================================================
\section{The Two-point Correlation Function}\label{sec:QSR}
%================================================================================
%================================================================================

In the framework of QCD sum rule~\cite{1979-Shifman-p385-447,
1985-Reinders-p1-1, 2000-Colangelo-p-}, we consider the two-point
correlation function:
\begin{eqnarray}
\nonumber \Pi_{\mu\nu,\rho\sigma}(q)&=& i\int
d^4xe^{iq\cdot x}\left\{\langle0|T[J_{\mu\nu}(x)J_{\rho\sigma}^{\dag}(y)]|0\rangle\right\}_{y\to 0}
\\
&=&\frac{1}{2}(\eta_{\mu\rho}\eta_{\nu\sigma}+\eta_{\mu\sigma}\eta_{\nu\rho}
-\frac{2}{3}\eta_{\mu\nu}\eta_{\rho\sigma})\Pi(q^2)+...,
 \label{equ:Pi}
\end{eqnarray}
with $\eta_{\mu\nu}=q_{\mu}q_{\nu}/q^2-g_{\mu\nu}$. The symbol $\left\{\cdots\right\}_{y\to 0}$  
means that we let $y=0$ after all the calculations except the Fourier transform. 
$J_{\mu\nu}$ is the tensor interpolating current with $J^{PC}=2^{--}$:
\begin{eqnarray}
  J_{\mu\nu}=\bar Q_1(x)(\gamma_{\mu}\gamma_5\overleftrightarrow{D_{\nu}}
  +\gamma_{\nu}\gamma_5\overleftrightarrow{D_{\mu}}-\frac{2}{3}\eta_{\mu\nu}\gamma_5
  \overleftrightarrow{D\!\!\!\slash})Q_2(x) . \label{current}
\end{eqnarray}
The $\eta_{\mu\nu}$ term is introduced to ensure the trace condition:
\begin{align}
  g^{\mu\nu}J_{\mu\nu}=0, \label{trace}
\end{align}
The covariant derivative $\overleftrightarrow{D_{\mu}}$ is defined
as:
\begin{align}
  \overleftrightarrow{D_{\mu}}&=\overrightarrow{D_{\mu}}-\overleftarrow{D_{\mu}},
  \\
  \overrightarrow{D_{\mu}}&=\overrightarrow{\partial_{\mu}}+ig\frac{\lambda^a}{2}A^a_{\mu},
  \overleftarrow{D_{\mu}}=\overleftarrow{\partial_{\mu}}-ig\frac{\lambda^a}{2}A^a_{\mu},
\end{align}

The interpolating current in Eq.~(\ref{current}) was first
constructed in Ref.~\cite{1982-Aliev-p401-401} to study the tensor
meson $f_2(1670)$ in the framework of QCD sum rule. It was also
introduced in Ref.~\cite{1985-Reinders-p1-1} as an operator with
$J^{PC}=2^{-+}$. However, we will show in Appendix~\ref{appendix1}
that it carries the odd $C$ parity through the charge conjugation
transformation. We also discuss this tensor operator in the
framework of the quark model. In fact, its $J_{ij}$ component
reduces to the D-wave tensor operator while its $J_{0i}$ component
reduces to the P-wave axial-vector operator in the
non-relativistic limit.

In order to study the tensor resonance, one should pick out the
intrinsic spin-2 tensor structure from the two-point correlation
function induced by the tensor current. In Eq.~(\ref{equ:Pi}),
$\Pi(q^2)$ contains contributions from the pure tensor states only
and ``..." represents the other possible structures from other
states such as the spin-1 states. At the hadron level, the
correlation function $\Pi(q^2)$ satisfies the dispersion relation:
\begin{eqnarray}
\Pi(q^2)=\int_{(m_1+m_2)^2}^{\infty}\frac{\rho(s)}{s-q^2-i\epsilon},
\label{Phenpi}
\end{eqnarray}
The hadronic spectral density $\rho(s)$ is usually assumed to take
the pole plus continuum contribution:
\begin{eqnarray}
\nonumber
\rho(s)&\equiv&\sum_n\delta(s-m_n^2)\langle0|\eta|n\rangle\langle n|\eta^+|0\rangle\\
&=&f_X^2m_X^6\delta(s-m_X^2)+ \mbox{continuum},   \label{Phenrho}
\end{eqnarray}
where $m_X$ denotes the mass of the resonance $X$ and $f_X$ stands
for the coupling constant of the tensor meson to the current
$J_{\mu\nu}$:
\begin{align}
  \langle0|J_{\mu\nu}|X\rangle=f_Xm_X^3\epsilon_{\mu\nu} \; .
\end{align}
Here $\epsilon_{\mu\nu}$ is the polarization tensor.

The correlation function can also be computed at the quark-gluon
level using the operator product expansion(OPE) method. For the
heavy quark systems, it's convenient to evaluate the Wilson
coefficient in the momentum space. In our calculation we only
consider the first order perturbative and various condensates
contributions, i.e., the bare quark loop in Fig.~\ref{fig1},
$\langle g_s^2G^2\rangle$ terms in Fig.~\ref{fig2} and $\langle
g_s^3fG^3\rangle$ terms in Fig.~\ref{fig3}. The massive quark
propagator in an external field $A_{\mu}(x)$ in the fixed-point
gauge is listed in Appendix~\ref{appendix2}. The quark lines
attached with gluon legs contain terms proportional to $y$. We
keep these terms throughout the evaluation and let $y$ go to zero
after finishing the derivatives. In Fig.~\ref{fig2} and
Fig.~\ref{fig3}, the diagrams with a gluon leg attached at the
right vertex are linear in $y$ and vanish after putting $y=0$.
They do not contribute to the correlation function.

%%%%%%%%%%%%%%%%%%%%%%%%%%%%%%%%%%%%%%%%%%%%%%%%%%%%%%%%%%%%%%%%%%%%%%%%%%%%%%
%\begin{figure}
\begin{center}
\begin{tabular}{l}
\scalebox{0.6}{\includegraphics{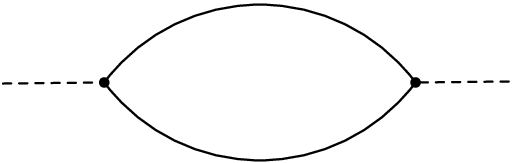}}
\end{tabular}
\figcaption{Feynman diagrams of the perturbative term for the
vacuum polarization. The solid and dashed lines denote the quark
line and interpolating current respectively.} \label{fig1}
\end{center}
%\end{figure}
%%%%%%%%%%%%%%%%%%%%%%%%%%%%%%%%%%%%%%%%%%%%%%%%%%%%%%%%%%%%%%%%%%%%%%%%%%%%%%
%%%%%%%%%%%%%%%%%%%%%%%%%%%%%%%%%%%%%%%%%%%%%%%%%%%%%%%%%%%%%%%%%%%%%%%%%%%%%%
%\begin{figure}
\begin{center}
\begin{tabular}{lllr}
\scalebox{0.4}{\includegraphics{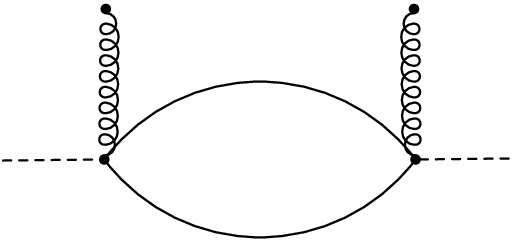}}&
\scalebox{0.4}{\includegraphics{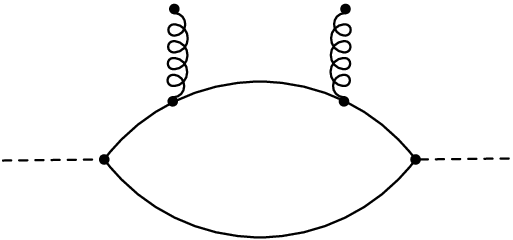}}&
\scalebox{0.4}{\includegraphics{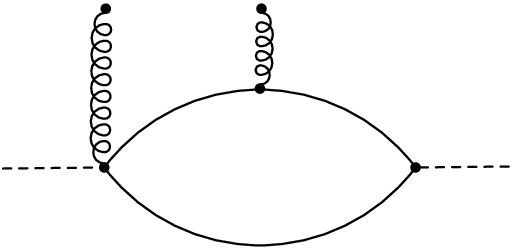}}&
\scalebox{0.4}{\includegraphics{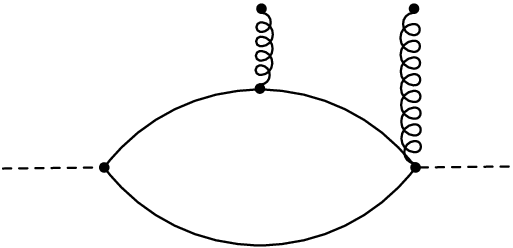}}\\
\scalebox{0.4}{\includegraphics{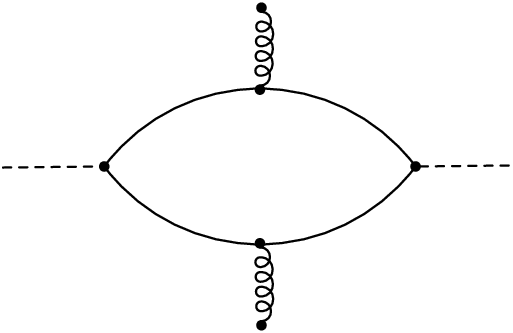}}&
\scalebox{0.4}{\includegraphics{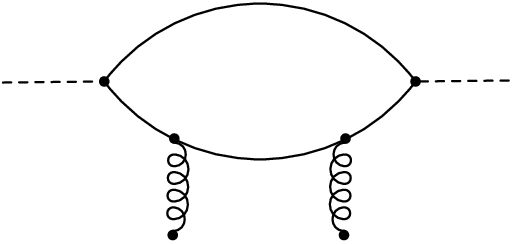}}&
\scalebox{0.4}{\includegraphics{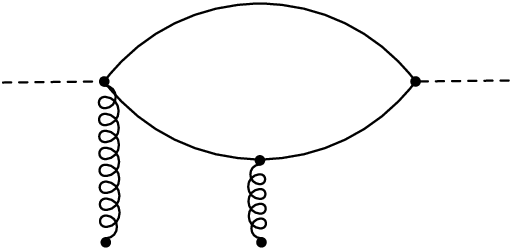}}&
\scalebox{0.4}{\includegraphics{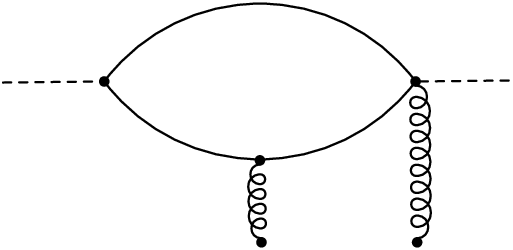}}
\end{tabular}
\figcaption{Feynman diagrams of the $\langle g_s^2G^2\rangle$
contribution to the vacuum polarization. The solid, curly and
dashed lines denote the quark, gluon lines and currents
respectively.} \label{fig2}
\end{center}
%\end{figure}
%%%%%%%%%%%%%%%%%%%%%%%%%%%%%%%%%%%%%%%%%%%%%%%%%%%%%%%%%%%%%%%%%%%%%%%%%%%%%%
%%%%%%%%%%%%%%%%%%%%%%%%%%%%%%%%%%%%%%%%%%%%%%%%%%%%%%%%%%%%%%%%%%%%%%%%%%%%%%
%\begin{figure}
\begin{center}
\begin{tabular}{lllr}
\scalebox{0.4}{\includegraphics{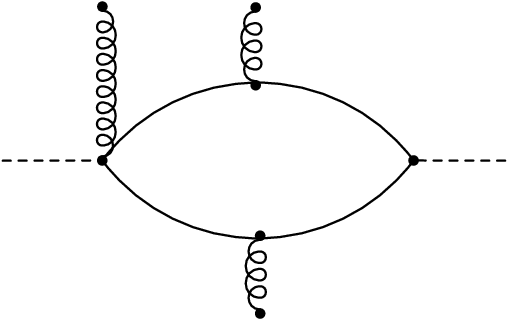}}&
\scalebox{0.4}{\includegraphics{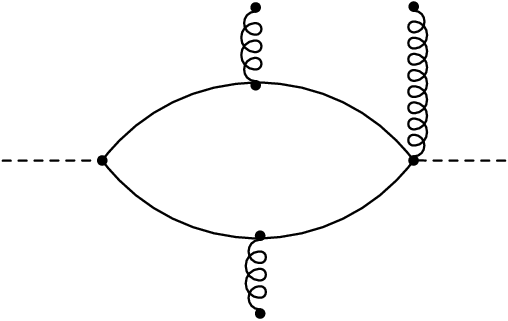}}&
\scalebox{0.4}{\includegraphics{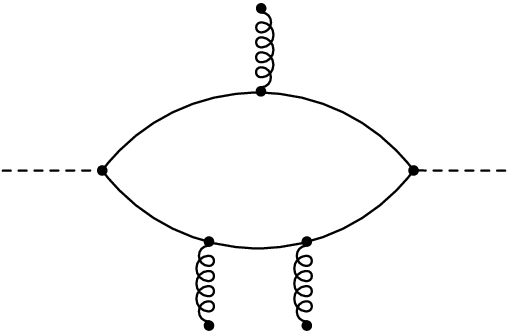}}&
\scalebox{0.4}{\includegraphics{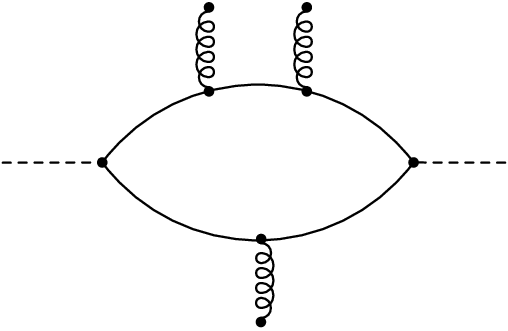}}\\
\scalebox{0.4}{\includegraphics{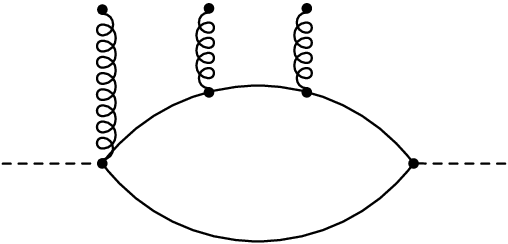}}&
\scalebox{0.4}{\includegraphics{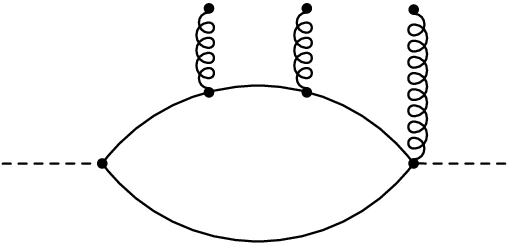}}&
\scalebox{0.4}{\includegraphics{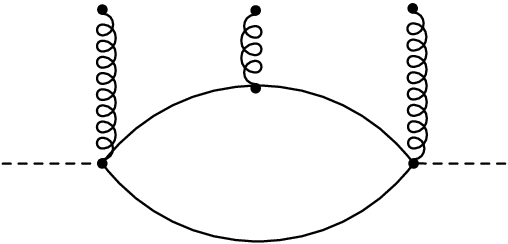}}&
\scalebox{0.4}{\includegraphics{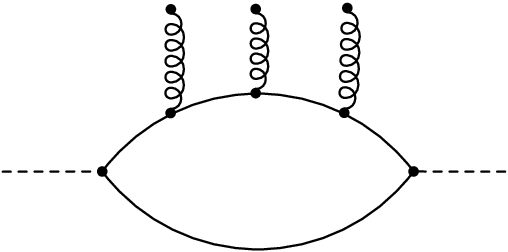}}\\
\scalebox{0.4}{\includegraphics{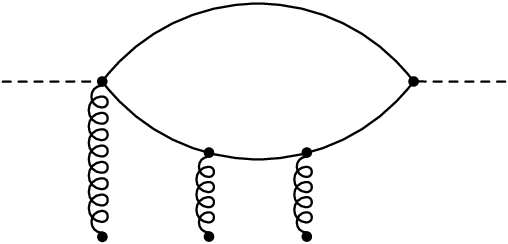}}&
\scalebox{0.4}{\includegraphics{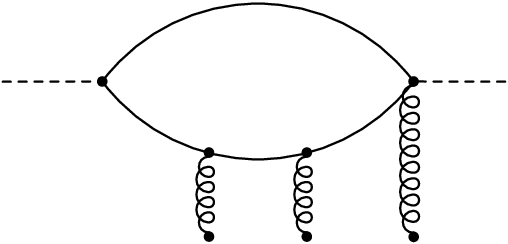}}&
\scalebox{0.4}{\includegraphics{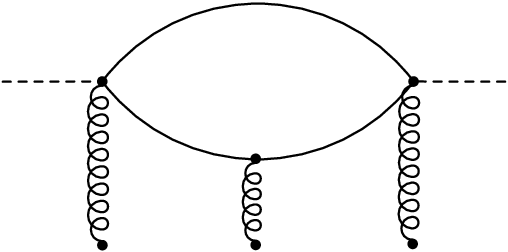}}&
\scalebox{0.4}{\includegraphics{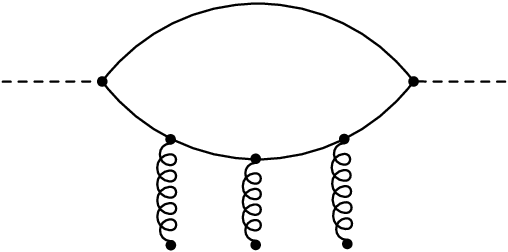}}
\end{tabular}
\figcaption{Feynman diagrams of the $\langle g_s^3fG^3\rangle$
contribution to the vacuum polarization.} \label{fig3}
\end{center}
%\end{figure}
%%%%%%%%%%%%%%%%%%%%%%%%%%%%%%%%%%%%%%%%%%%%%%%%%%%%%%%%%%%%%%%%%%%%%%%%%%%%%%
%%%%%%%%%%%%%%%%%%%%%%%%%%%%%%%%%%%%%%%%%%%%%%%%%%%%%%%%%%%%%%%%%%%%%%%%%%%%%%
%\begin{figure}
\begin{center}
\scalebox{0.4}{\includegraphics{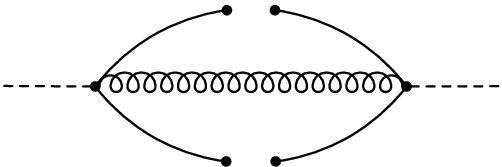}}\\
\begin{tabular}{lllr}
\scalebox{0.4}{\includegraphics{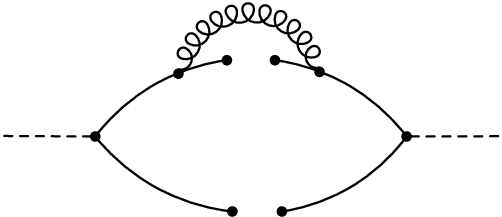}}&
\scalebox{0.4}{\includegraphics{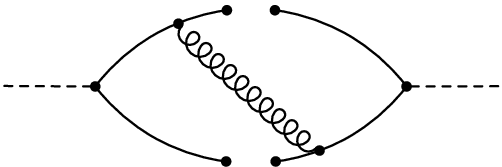}}&
\scalebox{0.4}{\includegraphics{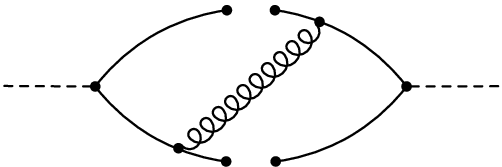}}&
\scalebox{0.4}{\includegraphics{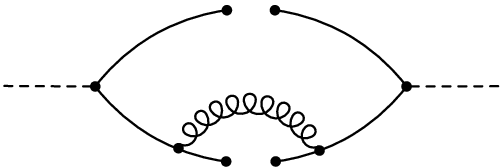}}\\
\scalebox{0.4}{\includegraphics{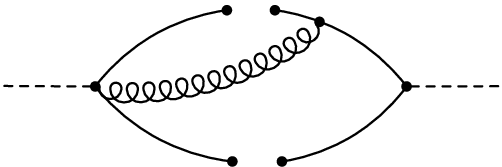}}&
\scalebox{0.4}{\includegraphics{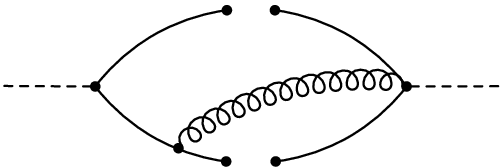}}&
\scalebox{0.4}{\includegraphics{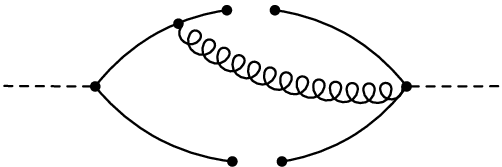}}&
\scalebox{0.4}{\includegraphics{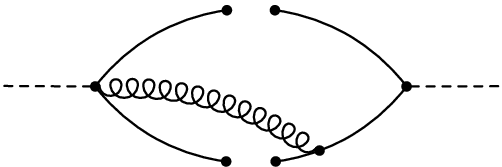}}
\end{tabular}
\figcaption{Feynman diagrams of the $\langle g_s\bar qq\rangle^2$
contribution to the vacuum polarization.} \label{fig4}
\end{center}
%\end{figure}
%%%%%%%%%%%%%%%%%%%%%%%%%%%%%%%%%%%%%%%%%%%%%%%%%%%%%%%%%%%%%%%%%%%%%%%%%%%%%%

In order to suppress the higher states contributions, it is
significant to perform the Borel transform to the correlation
function, which also helps improve the convergence of the OPE
series. With the assumption of the quark-hadron duality, we derive
the tensor meson sum rule:
\begin{eqnarray}
\Pi(M_B^2)=f_X^2m_X^6e^{-m_X^2/M_B^2}=\int_{(m_1+m_2)^2}^{s_0}dse^{-s/M_B^2}\rho(s),
\label{sumrule}
\end{eqnarray}
where $s_0$ is the threshold parameter. Then we can extract the
meson mass $m_X$:
\begin{eqnarray}
m_X^2=-\frac{\frac{\partial}{\partial(1/M_B^2)}\Pi(M_B^2)}{\Pi(M_B^2)}=\frac{\int_{(m_1+m_2)^2}^{s_0}dse^{-s/M_B^2}s\rho(s)}
{\int_{(m_1+m_2)^2}^{s_0}dse^{-s/M_B^2}\rho(s)}. \label{mass}
\end{eqnarray}
After performing the Borel transform, the correlation
function reads:
\begin{align}
\nonumber
\Pi^{pert}(M_B^2)&=\frac{6}{\pi^2}\int_{(m_1+m_2)^2}^{s_0}dse^{-s/M_B^2}\int_{x_{min}}^{x_{max}}dx
\Big\{\big[m_1^2-s(1-x)\big]x+m_2^2(1-x)\Big\}\big[m_1m_2-(m_1^2-m_2^2+sx)x\big],
\non
\Pi^{\GGa}_1(M_B^2)&=\frac{\GGb}{24\pi^2}\int_{(m_1+m_2)^2}^{s_0}dse^{-s/M_B^2}\int_{x_{min}}^{x_{max}}dx
(8x^2-4x-5), \non
\Pi^{\GGa}_2(M_B^2)&=-\frac{\GGb}{24\pi^2}\int_0^1dx\bigg[\frac{(4x^3-2x^2-9x+2)m_1^2}{(1-x)^2}
+\frac{2(9x^2-9x+4)m_1m_2}{(1-x)x}-\frac{(4x^2-4x-5)m_2^2}{x}
\non&\hspace{3.4cm}+\frac{2x(x+2)m_1^4}{M_B^2(1-x)^3}
-\frac{(2x+3)m_1^3m_2}{M_B^2(1-x)^2}+\frac{2(x^2-x+1)m_1^2m_2^2}{M_B^2(1-x)^2x}
\non&\hspace{3.4cm}+\frac{(2x-3)m_1m_2^3}{M_B^2 x^2}-\frac{2
m_2^4}{M_B^2 x^2}\bigg]e^{-\tilde{m}^2/M_B^2}, \non
\Pi^{\GGG}(M_B^2)&=\frac{\langle
g_s^3fGGG\rangle}{288\pi^2}\int_0^1dx\bigg[\frac{(20x^3+3x^2-66x+35)}{(1-x)^2}
-\frac{3(2x^2+3x-2)m_1^4}{M_B^4(1-x)^4}+\frac{2(5x^2+4x-2)m_1^3m_2}{M_B^4(1-x)^3x}
\non&\hspace{3.4cm}+\frac{(10x^3-32x^2+39x-14)m_1^2m_2^2}
{M_B^4(1-x)^3x^2}-\frac{(32x^4-71x^3+14x^2+6x-7)m_1^2}{M_B^2(1-x)^3x}
\non&\hspace{3.4cm}+\frac{6(x^2-4x+2)m_1m_2^3}{M_B^4(1-x)x^3}
-\frac{(32x^4-46x^3+124x^2-96x+33)m_1m_2}{M_B^2(1-x)^2x^2}
\non&\hspace{3.4cm}-\frac{3m_2^4}{M_B^4(1-x)x^2}
+\frac{(32x^3-87x^2+72x-20)m_2^2}{M_B^2(1-x)^2x}\bigg]e^{-\tilde{m}^2/M_B^2},
\non \label{correlation} \Pi^{\qGq}(M_B^2)&=-2m_2\langle \bar
q_1g_s\sigma\cdot Gq_1\rangle e^{-m_2^2/M_B^2} -2m_1\langle \bar
q_2g_s\sigma\cdot Gq_2\rangle e^{-m_1^2/M_B^2},
\\
\Pi^{\qqb}(M_B^2)&=-\frac{16}{9}g_s^2\langle\bar
q_1q_1\rangle\langle\bar q_2q_2\rangle.
\end{align}
where
$x_{min}=\frac{1}{2}\Big\{1+\frac{m_2^2-m_1^2}{s}-\big[(1+\frac{m_2^2-m_1^2}{s})^2-4m_2^2/s\big]^{1/2}\Big\},
x_{max}
=\frac{1}{2}\Big\{1+\frac{m_2^2-m_1^2}{s}+\big[(1+\frac{m_2^2-m_1^2}{s})^2-4m_2^2/s\big]^{1/2}\Big\}$
and $\tilde{m}^2= \frac{xm_1^2+m_2^2(1-x)}{(1-x)x}$. Only the
two-gluon condensate contribution $\Pi^{\GGa}(M_B^2)$ and
tri-gluon condensate contribution $\Pi^{\GGG}(M_B^2)$ are involved
for the heavy quark systems ($Q_1,Q_2=c,b$). For the $\bar qq,
\bar qs, \bar ss, \bar qc, \bar sc, \bar qb$ and $\bar sb$
systems, the four quark condensate $g_s^2\langle\bar
q_1q_1\rangle\langle\bar q_2q_2\rangle$ (only for the light quarks
systems, $q_1,q_2=u,d,s$) and the quark-gluon mixed condensate
$\langle \bar qg_s\sigma\cdot Gq\rangle$ are also needed. Among
these systems, the charge neutral ones have the quantum numbers
$J^{PC}=2^{--}$ and the charged ones have $J^{P}=2^{-}$ with no
definite $C$ parity. The quark condensate $\qq$ does not
contribute to the intrinsic tensor structure. The four quark
condensate $g_s^2\langle\bar q_1q_1\rangle\langle\bar
q_2q_2\rangle$ as shown in Fig.~\ref{fig4}, always plays an
important role in the conventional $q\bar q$ meson sum
rules~\cite{1979-Shifman-p385-447, 1985-Reinders-p1-1}. In the
present case, only the first diagram in Fig.~\ref{fig4} contributes
to the correlation function. All the other diagrams vanish due to
the special Lorentz structure of the current.

%================================================================================
%================================================================================
\section{Numerical Analysis}\label{sec:numerical}
%================================================================================
%================================================================================

In the QCD sum rule analysis we use the following values of the
quark masses and various condensates~\cite{1979-Shifman-p385-447,
1985-Reinders-p1-1, 2010-Nakamura-p75021-75021,
2002-Jamin-p237-243}:
\begin{eqnarray}
\nonumber &&m_u(2\text{ GeV})=(2.9\pm0.6)\text{ MeV} \, , \non
&&m_d(2\text{ GeV})=(5.2\pm0.9)\text{ MeV} \, , \non &&m_q(2\text{
GeV})=(4.0\pm0.7)\text{ MeV} \, , \non &&m_s(2\text{
GeV})=(101^{+29}_{-21})\text{ MeV} \, , \non
&&\qq=-(0.23\pm0.03)^3\text{ GeV}^3 \, , \non &&\qGq=-M_0^2\qq\, ,
\non &&M_0^2=(0.8\pm0.2)\text{ GeV}^2 \, ,
\\ \label{parameters}
&&\langle\bar ss\rangle/\qq=0.8\pm0.2 \, , \non &&\GGb=(0.88\pm0.14)\text{
GeV}^4 \, , \non &&\GGGb=(0.087\pm0.011)\text{ GeV}^6 \, .
\end{eqnarray}
where the $up, down$ and $strange$ quark masses are the current
quark masses in a mass-independent subtraction scheme such as
$\overline{MS}$ at a scale $\mu=2$ GeV. The running charm quark
mass has been determined by the moment sum rule in
Refs.~\cite{1994-Dominguez-p184-189, 1994-Narison-p73-83,
2001-Eidemuller-p203-210, 2001-Kuhn-p588-602,
2003-Ioffe-p229-241}. Recently, the value has been updated by
using the four loop results for the vacuum polarization
function~\cite{2009-Chetyrkin-p74010-74010}.

Since we have not calculated the next leading order radiative
correction due to the complicated interpolating current in the
present work, it is desirable to extract the charm quark mass
within the same QCD sum rule formalism using the experimental
$J/\psi$ mass as input. The $J/\psi$ sum rule derived from the
interpolating current $j_{\mu}=\bar c\gamma_{\mu}c$ was known very
well \cite{1985-Reinders-p1-1}. With the same criterion of the
present tensor current and keeping only the leading order
perturbative term, gluon condensate and tri-gluon condensate
contributions, we show the Borel sum rule results in
Fig.~\ref{fig5} using the experimental
data~\cite{2010-Nakamura-p75021-75021}. The extracted charm  and
bottom quark mass are $m_c=(1.35\pm0.08) $ GeV and $m_b=(4.60\pm0.18)$ GeV 
as shown in Fig. \ref{fig5}.

%%%%%%%%%%%%%%%%%%%%%%%%%%%%%%%%%%%%%%%%%%%%%%%%%%%%%%%%%%%%%%%%%%%%%%%%%%%%%%
%\begin{figure}
\begin{center}
\begin{tabular}{lr}
\scalebox{0.60}{\includegraphics{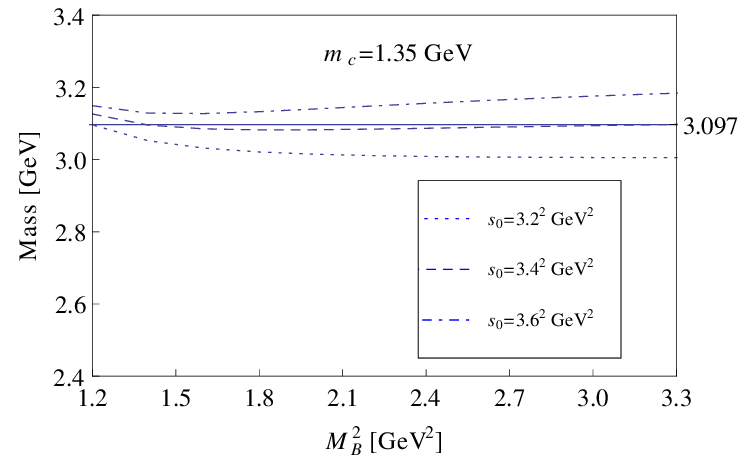}}&
\scalebox{0.60}{\includegraphics{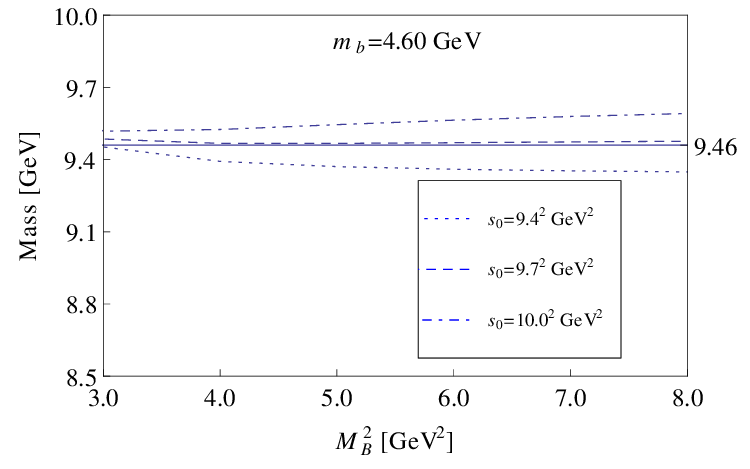}}
\end{tabular}
\figcaption{Running masses of the charm and bottom quarks in the
Borel sum rules for $J/\psi$ and $\Upsilon(1S)$, respectively.} \label{fig5}
\end{center}
%\end{figure}
%%%%%%%%%%%%%%%%%%%%%%%%%%%%%%%%%%%%%%%%%%%%%%%%%%%%%%%%%%%%%%%%%%%%%%%%%%%%%%

After performing the Borel transform, the correlation
function in Eq.~(\ref{correlation}) is the function of the threshold
parameter $s_0$ and Borel mass $M_B$. In the Borel sum rules
analysis, there should exist suitable working regions of these two
parameters in order to obtain a stable mass sum rules. For this
purpose we choose the value of $s_0$ around which the extracted
mass $m_X$ is stable with the variation of $M_B^2$. In
Eq.~(\ref{correlation}), the exponential weight functions suppress
the higher states contributions for the small value of $M_B^2$.
However, the convergence of the OPE series becomes bad if $M_B^2$
is too small. These two opposite requirements restrict the domain
of the Borel mass. We define the pole contribution (PC) as:
\begin{eqnarray}
\text{PC}=\frac{\int_{4m_c^2}^{s_0}dse^{-s/M_B^2}\rho(s)}{\int_{4m_c^2}^{\infty}dse^{-s/M_B^2}\rho(s)},
\label{pc}
\end{eqnarray}
The constraint of a relatively large pole contribution leads to
the upper bound $M^2_{max}$ of the Borel parameter while the
requirement of the OPE convergence yields the lower bound
$M^2_{min}$.

Using the parameter values in Eq.~(\ref{parameters}), we study the
tensor charmonium system by considering only $\Pi^{pert}(M_B^2)$,
$\Pi^{\GGa}(M_B^2)$ and $\Pi^{\GGG}(M_B^2)$ in
Eq.~(\ref{correlation}) with $m_1=m_2=m_c$. The two-gluon condensate
contribution is the dominant correction to the correlation
function in this situation. The lower bound of the Borel parameter
is obtained as $M^2_{min}=2.7$ GeV$^2$ by requiring that the
two-gluon condensate correction be less than one fifth of the
perturbative term and the tri-gluon condensate correction less
than one fifth of the two-gluon condensate correction. The upper
limit of $M_B^2$ is the function of $s_0$ as shown in
Eq.~(\ref{pc}). We choose $\sqrt{s_0}=4.4$ GeV around which the
variation of the extracted mass $m_X$ with $M_B^2$ is the minimum,
as shown in Fig.~\ref{fig6}a. Then we require that the pole
contribution be larger than $50\%$ to get the upper bound of the
Borel mass $M^2_{max}=4.0$ GeV$^2$.

By performing the numerical analysis in the domain $2.7\leq
M_B^2\leq4.0$ GeV$^2$, we obtain a stable mass sum rule of the
$2^{--}$ charmonium system. The dependence of the extracted mass
$m_X$ with the Borel parameter is very weak in this domain of
$M_B^2$, as shown in Fig.~\ref{fig6}b. The extracted mass for the
possible charmonium tensor state is about $3.97\pm0.25$ GeV. As
expected naively in the quark model, this value is slightly higher
than the mass of the lowest D-wave $1^{--}$ charmonium state
$\psi(3770)$.

We extend the same analysis to the $\bar cb$ and $\bar bb$ heavy
quark systems and collect the numerical results in Table~\ref{table1}.
The masses of the $B_c$ and bottomonium tensor states are
extracted to be $7.08\pm0.34$ and $10.13\pm0.34$ GeV, respectively.
The errors are from the uncertainties of the quark masses, QCD condensates, 
the threshold values and the Borel parameter. 
The mass of the $2^-$ $\bar cb$ state was predicted to be around $7.0-7.1$ GeV
in Refs.~\cite{1981-Eichten-p2724-2724, 1988-Gershtein-p327-327,
1993-Chen-p350-350, 1994-Eichten-p5845-5856}, which is consistent
with our calculation.

%%%%%%%%%%%%%%%%%%%%%%%%%%%%%%%%%%%%%%%%%%%%%%%%%%%%%%%%%%%%%%%%%%%%%%%%%%%%%%
%\begin{figure}
\begin{center}
\begin{tabular}{lr}
\scalebox{0.6}{\includegraphics{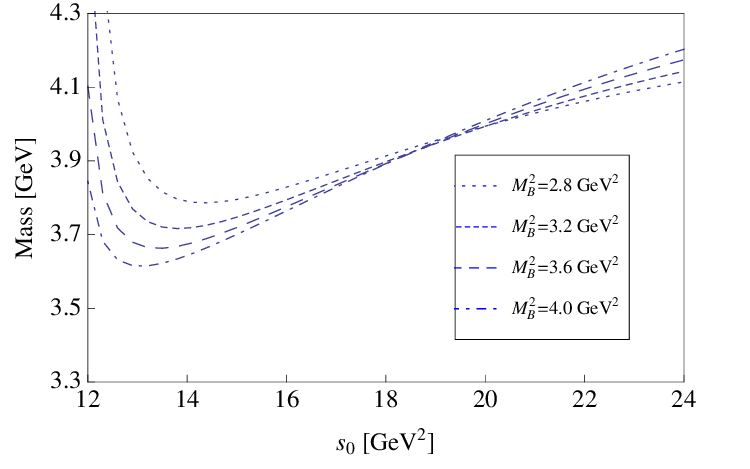}}&
\scalebox{0.6}{\includegraphics{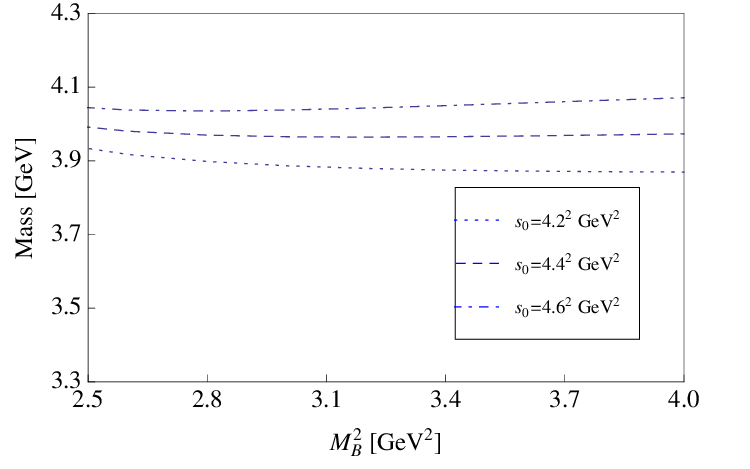}}
\end{tabular}
\centerline{\hspace{0.05in} {(a)} \hspace{3in}{ (b)}}
\figcaption{Variation of $m_X$ with $s_0$(a) and $M^2_B$(b)
for the $2^{--}$ charmonium state in the framework of the Borel
sum rules.} \label{fig6}
\end{center}
%\end{figure}
%%%%%%%%%%%%%%%%%%%%%%%%%%%%%%%%%%%%%%%%%%%%%%%%%%%%%%%%%%%%%%%%%%%%%%%%%%%%%%
%%%%%%%%%%%%%%%%%%%%%%%%%%%%%%%%%%%%%%%%%%%%%%%%%%%%%%%%%%%%%%%%%%%%%%%%%%%%%%
%\begin{figure}
\begin{center}
\begin{tabular}{lr}
\scalebox{0.6}{\includegraphics{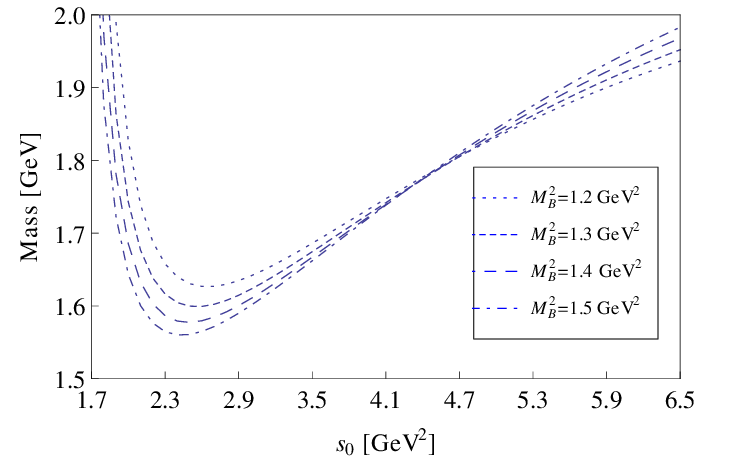}}&
\scalebox{0.6}{\includegraphics{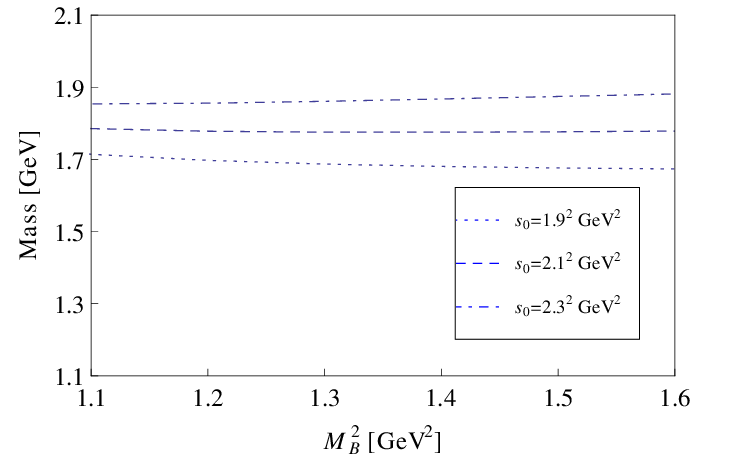}}
\end{tabular}
\figcaption{Variation of $m_X$ with $s_0$ and $M^2_B$
for the $2^{--}$ $\bar qq$ state in the framework of the Borel
sum rules.} \label{fig7}
\end{center}
%\end{figure}
%%%%%%%%%%%%%%%%%%%%%%%%%%%%%%%%%%%%%%%%%%%%%%%%%%%%%%%%%%%%%%%%%%%%%%%%%%%%%%
%%%%%%%%%%%%%%%%%%%%%%%%%%%%%%%%%%%%%%%%%%%%%%%%%%%%%%%%%%%%%%%%%%%%%%%%%%%%%%
%\begin{figure}
\begin{center}
\begin{tabular}{lr}
\scalebox{0.6}{\includegraphics{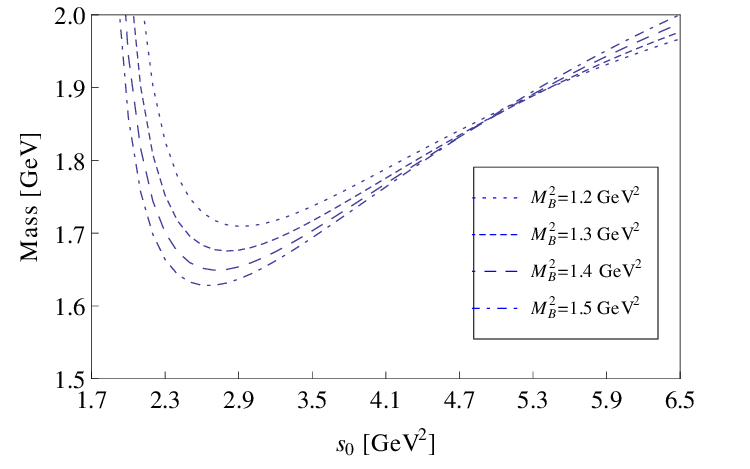}}&
\scalebox{0.6}{\includegraphics{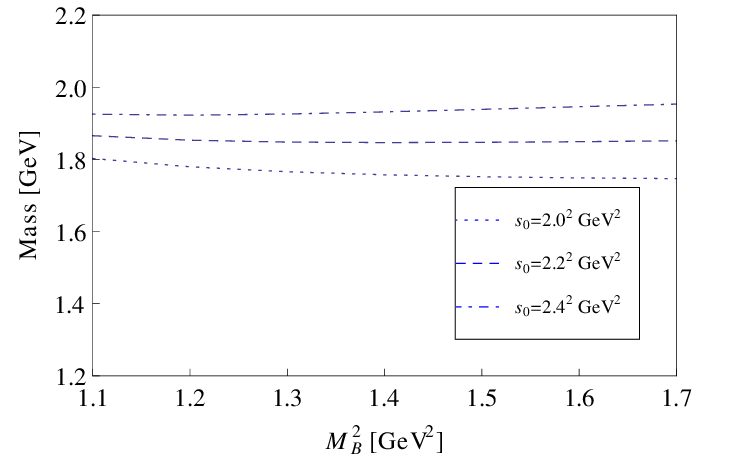}}
\end{tabular}
\figcaption{Variation of $m_X$ with $s_0$ and $M^2_B$
for the $2^{-}$ $\bar qs$ state in the framework of the Borel
sum rules.} \label{fig8}
\end{center}
%\end{figure}
%%%%%%%%%%%%%%%%%%%%%%%%%%%%%%%%%%%%%%%%%%%%%%%%%%%%%%%%%%%%%%%%%%%%%%%%%%%%%%
%%%%%%%%%%%%%%%%%%%%%%%%%%%%%%%%%%%%%%%%%%%%%%%%%%%%%%%%%%%%%%%%%%%%%%%%%%%%%%
%\begin{figure}
\begin{center}
\begin{tabular}{lr}
\scalebox{0.6}{\includegraphics{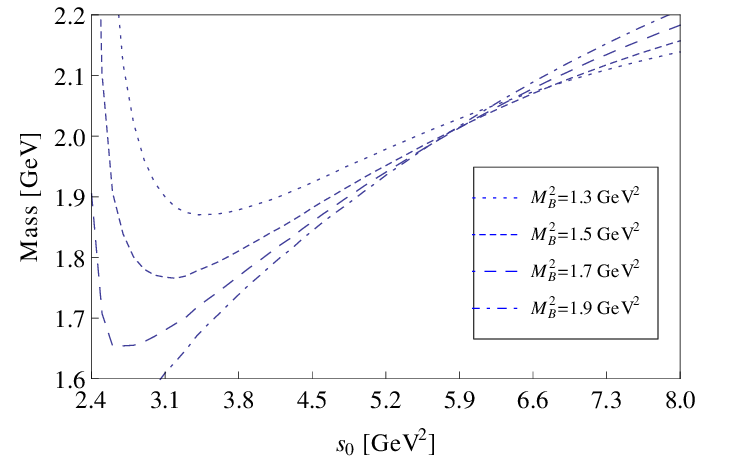}}&
\scalebox{0.6}{\includegraphics{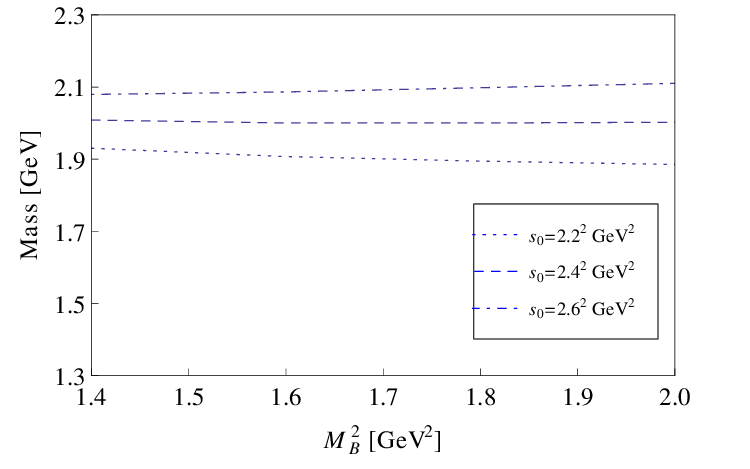}}
\end{tabular}
\figcaption{Variation of $m_X$ with $s_0$ and $M^2_B$
for the $2^{--}$ $\bar ss$ state in the framework of the Borel
sum rules.} \label{fig9}
\end{center}
%\end{figure}
%%%%%%%%%%%%%%%%%%%%%%%%%%%%%%%%%%%%%%%%%%%%%%%%%%%%%%%%%%%%%%%%%%%%%%%%%%%%%%
%%%%%%%%%%%%%%%%%%%%%%%%%%%%%%%%%%%%%%%%%%%%%%%%%%%%%%%%%%%%%%%%%%%%%%%%%%%%%%
%\begin{figure}
\begin{center}
\begin{tabular}{lr}
\scalebox{0.6}{\includegraphics{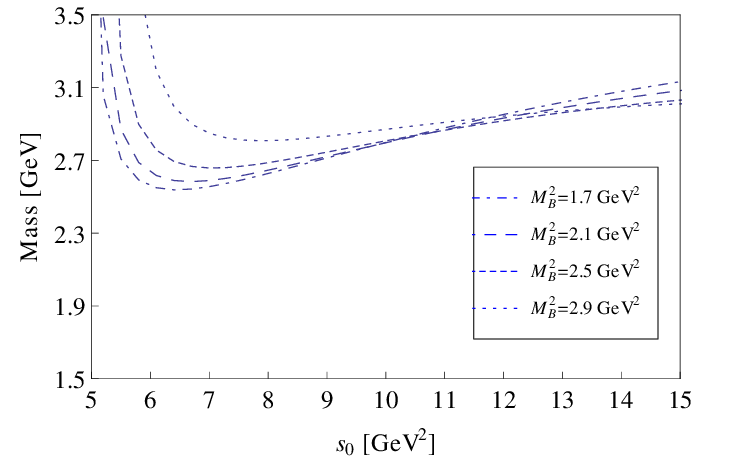}}&
\scalebox{0.6}{\includegraphics{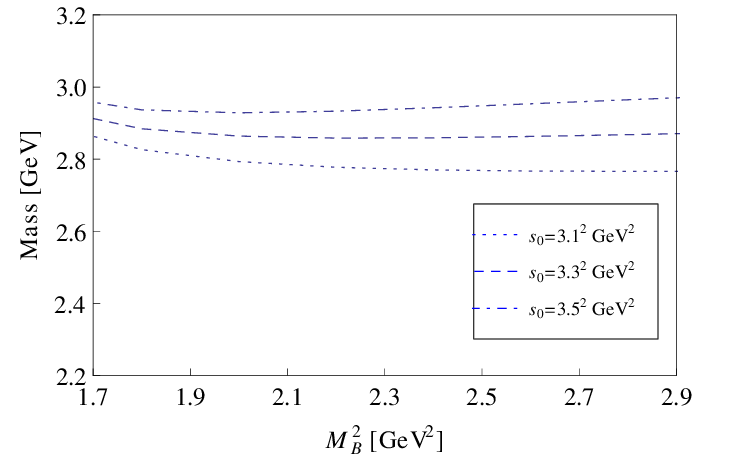}}
\end{tabular}
\figcaption{Variation of $m_X$ with $s_0$ and $M^2_B$
for the $2^{-}$ $\bar qc$ state in the framework of the Borel
sum rules.} \label{fig10}
\end{center}
%\end{figure}
%%%%%%%%%%%%%%%%%%%%%%%%%%%%%%%%%%%%%%%%%%%%%%%%%%%%%%%%%%%%%%%%%%%%%%%%%%%%%%
%%%%%%%%%%%%%%%%%%%%%%%%%%%%%%%%%%%%%%%%%%%%%%%%%%%%%%%%%%%%%%%%%%%%%%%%%%%%%%
%\begin{figure}
\begin{center}
\begin{tabular}{lr}
\scalebox{0.6}{\includegraphics{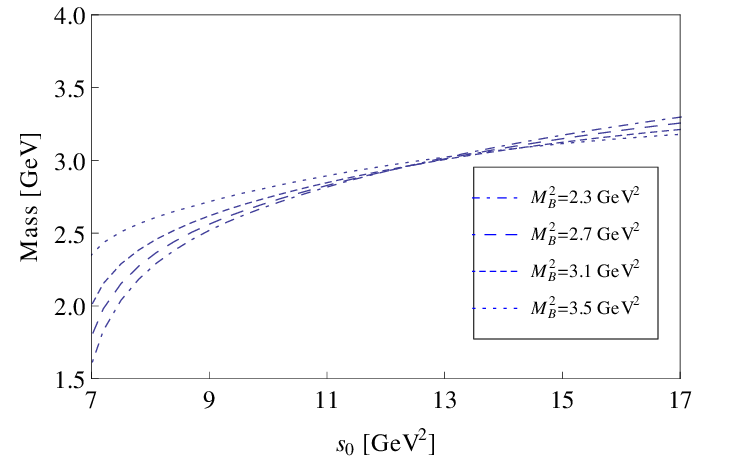}}&
\scalebox{0.6}{\includegraphics{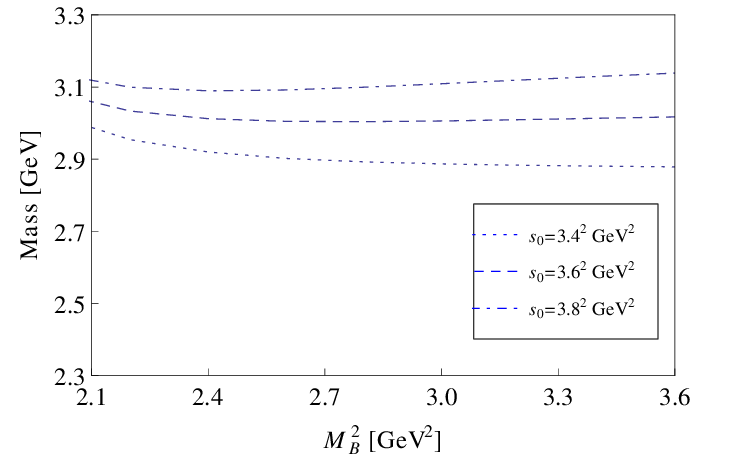}}
\end{tabular}
\figcaption{Variation of $m_X$ with $s_0$ and $M^2_B$
for the $2^{-}$ $\bar sc$ state in the framework of the Borel
sum rules.} \label{fig11}
\end{center}
%\end{figure}
%%%%%%%%%%%%%%%%%%%%%%%%%%%%%%%%%%%%%%%%%%%%%%%%%%%%%%%%%%%%%%%%%%%%%%%%%%%%%%
%%%%%%%%%%%%%%%%%%%%%%%%%%%%%%%%%%%%%%%%%%%%%%%%%%%%%%%%%%%%%%%%%%%%%%%%%%%%%%
%\begin{figure}
\begin{center}
\begin{tabular}{lr}
\scalebox{0.6}{\includegraphics{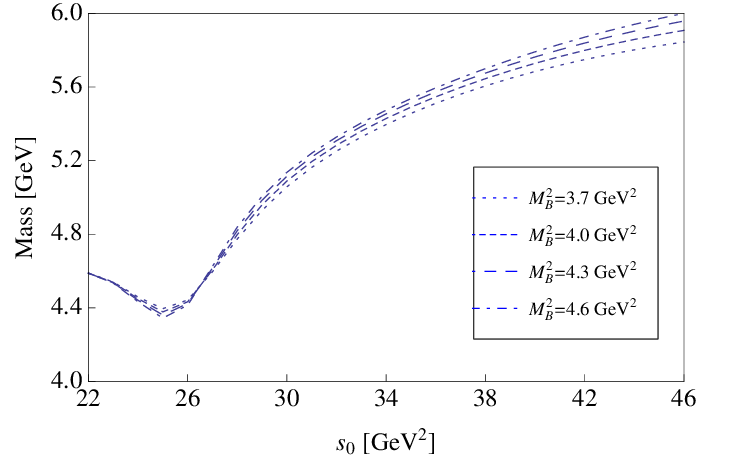}}&
\scalebox{0.6}{\includegraphics{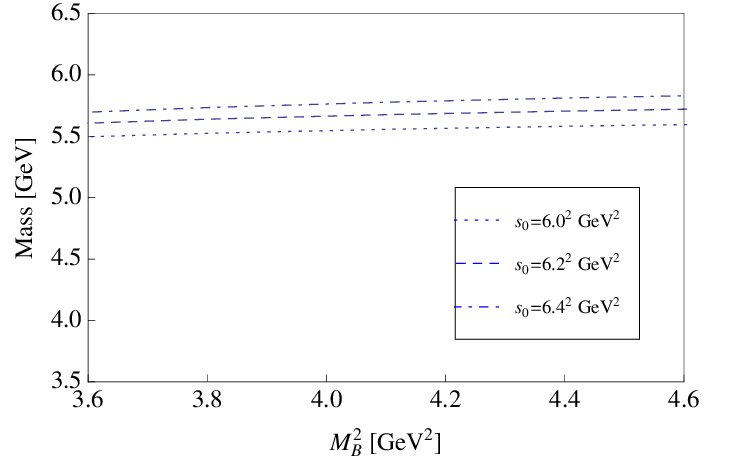}}
\end{tabular}
\figcaption{Variation of $m_X$ with $s_0$ and $M^2_B$
for the $2^{-}$ $\bar qb$ state in the framework of the Borel
sum rules.} \label{fig12}
\end{center}
%\end{figure}
%%%%%%%%%%%%%%%%%%%%%%%%%%%%%%%%%%%%%%%%%%%%%%%%%%%%%%%%%%%%%%%%%%%%%%%%%%%%%%
%%%%%%%%%%%%%%%%%%%%%%%%%%%%%%%%%%%%%%%%%%%%%%%%%%%%%%%%%%%%%%%%%%%%%%%%%%%%%%
%\begin{figure}
\begin{center}
\begin{tabular}{lr}
\scalebox{0.6}{\includegraphics{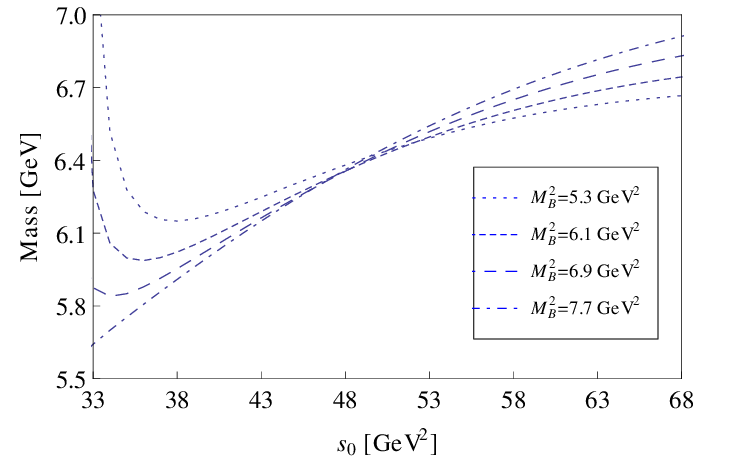}}&
\scalebox{0.6}{\includegraphics{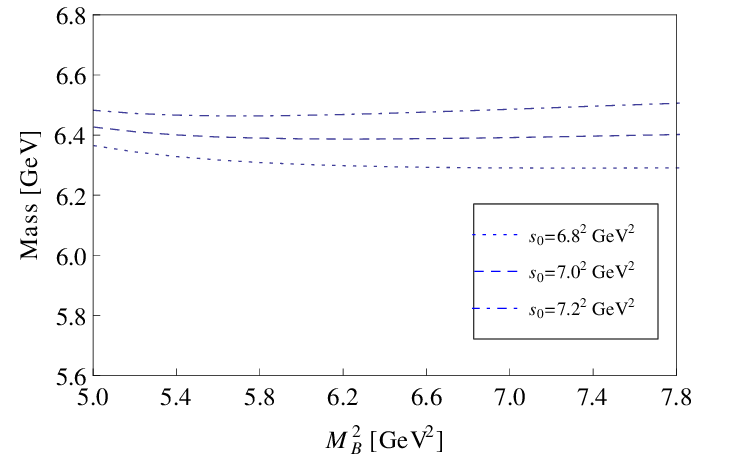}}
\end{tabular}
\figcaption{Variation of $m_X$ with $s_0$ and $M^2_B$
for the $2^{-}$ $\bar sb$ state in the framework of the Borel
sum rules.} \label{fig13}
\end{center}
%\end{figure}
%%%%%%%%%%%%%%%%%%%%%%%%%%%%%%%%%%%%%%%%%%%%%%%%%%%%%%%%%%%%%%%%%%%%%%%%%%%%%%
%%%%%%%%%%%%%%%%%%%%%%%%%%%%%%%%%%%%%%%%%%%%%%%%%%%%%%%%%%%%%%%%%%%%%%%%%%%%%%
%\begin{figure}
\begin{center}
\begin{tabular}{lr}
\scalebox{0.6}{\includegraphics{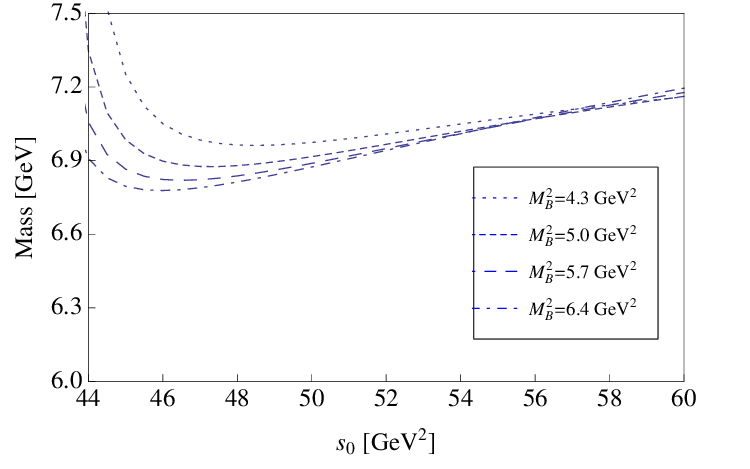}}&
\scalebox{0.6}{\includegraphics{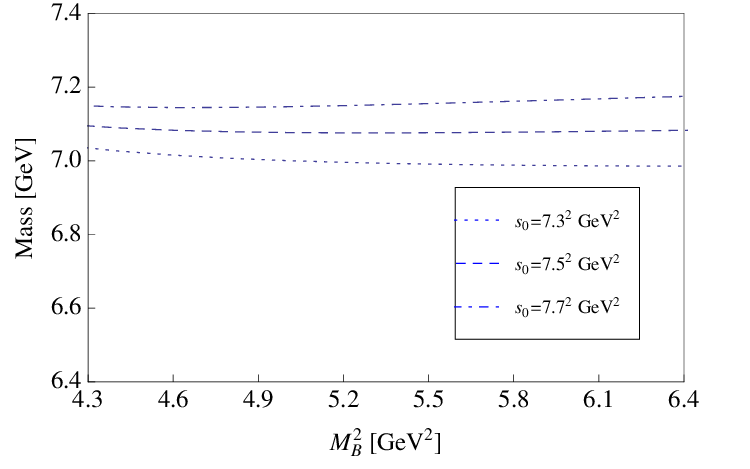}}
\end{tabular}
\figcaption{Variation of $m_X$ with $s_0$ and $M^2_B$
for the $2^{-}$ $\bar cb$ state in the framework of the Borel
sum rules.} \label{fig14}
\end{center}
%\end{figure}
%%%%%%%%%%%%%%%%%%%%%%%%%%%%%%%%%%%%%%%%%%%%%%%%%%%%%%%%%%%%%%%%%%%%%%%%%%%%%%
%%%%%%%%%%%%%%%%%%%%%%%%%%%%%%%%%%%%%%%%%%%%%%%%%%%%%%%%%%%%%%%%%%%%%%%%%%%%%%
%\begin{figure}
\begin{center}
\begin{tabular}{lr}
\scalebox{0.6}{\includegraphics{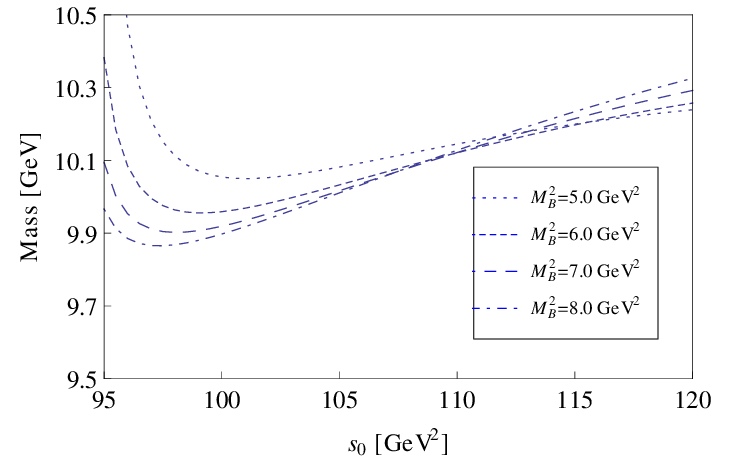}}&
\scalebox{0.6}{\includegraphics{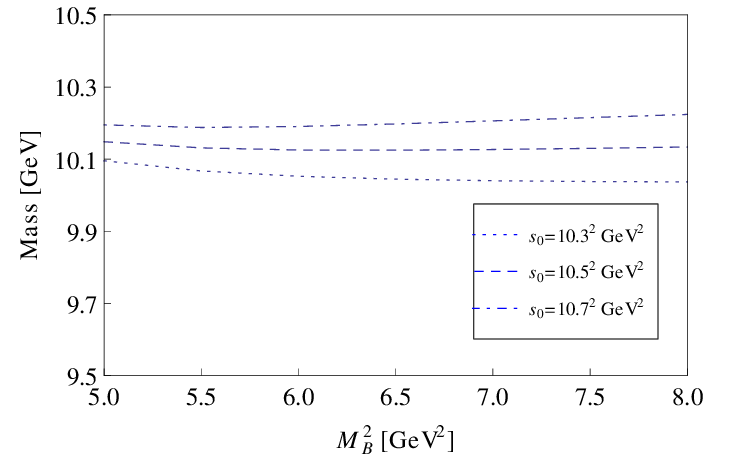}}
\end{tabular}
\figcaption{Variation of $m_X$ with $s_0$ and $M^2_B$
for the $2^{--}$ bottomonium state in the framework of the Borel
sum rules.} \label{fig15}
\end{center}
%\end{figure}
%%%%%%%%%%%%%%%%%%%%%%%%%%%%%%%%%%%%%%%%%%%%%%%%%%%%%%%%%%%%%%%%%%%%%%%%%%%%%%

We also extend the analysis to the $\bar qq, \bar qs, \bar ss,
\bar qc, \bar sc, \bar qb$ and $\bar sb$ systems, where $q$
denotes $u$ or $d$ quark. As mentioned in Sec.~\ref{sec:QSR}, the
four quark condensate $g_s^2\langle\bar q_1q_1\rangle\langle\bar
q_2q_2\rangle$ and the quark-gluon mixed condensate $\langle \bar
qg_s\sigma\cdot Gq\rangle$ should be considered now. The
corresponding parameters such as the quark masses and the
condensates should be used for different systems in
Eq.~(\ref{correlation}). Especially for the $\bar qc$ ($\bar sc$)
system, only the light (strange) quark-gluon mixed condensate
needs to be considered. After performing the numerical analysis,
the variations of the mass with the threshold value $s_0$ and
Borel parameter $M_B^2$ are shown in Figs.~\ref{fig6}-\ref{fig15}.
We show the Borel window, the threshold value and the extracted
mass for all these systems in Table~\ref{table1}. As shown in
Table~\ref{table1}, the light tensor meson mass is about 1.78 GeV,
which is consistent with the value 1.63 GeV in
Ref.~\cite{1982-Aliev-p401-401}.

%
%%%%%%%%%%%%%%%%%%%%%%%%%%%%%%%%%%%%%%%%%%%%%%%%%%%%%%%%%%%%%%%%%%%%%%%%%%%%%%%%%%%%%%%%%%%%%%%%%%%%%%%%%%%%%%%%%%%%%%%%%%%%%%
\begin{center}
\begin{tabular}{cccc}
\hlinewd{.8pt}
System           & $s_0(\mbox{GeV}^2)$  & $[M^2_{\mbox{min}}$,$M^2_{\mbox{max}}](\mbox{GeV}^2)$ & $m_X$\mbox{(GeV)} \\
\hline
$\bar qq$        &  $2.1^2$             & $1.2 - 1.5$                                          & $1.78\pm0.12$     \\
$\bar qs$        &  $2.2^2$             & $1.2 - 1.6$                                          & $1.85\pm0.14$     \\
$\bar ss$        &  $2.4^2$             & $1.4 - 1.9$                                          & $2.00\pm0.16$     \\
$\bar qc$        &  $3.3^2$             & $1.6 - 2.9$                                          & $2.86\pm0.14$     \\
$\bar sc$        &  $3.6^2$             & $2.0 - 3.5$                                          & $3.01\pm0.21$     \\
$\bar cc$        &  $4.4^2$             & $2.7 - 4.0$                                          & $3.97\pm0.25$     \\
$\bar qb$        &  $6.2^2$             & $3.7 - 4.6$                                          & $5.66\pm0.33$     \\
$\bar sb$        &  $7.0^2$             & $4.1 - 7.7$                                          & $6.40\pm0.25$     \\
$\bar cb$        &  $7.5^2$             & $4.3 - 6.4$                                          & $7.08\pm0.34$     \\
$\bar bb$        &  $10.5^2$            & $5.0 - 8.0$                                          & $10.13\pm0.34$    \\
\hlinewd{.8pt}
\end{tabular}
\tabcaption{Numerical results for the various $2^{-(-)}$ tensor states in the framework of the
Borel sum rules. \label{table1}}
\end{center}
%%%%%%%%%%%%%%%%%%%%%%%%%%%%%%%%%%%%%%%%%%%%%%%%%%%%%%%%%%%%%%%%%%%%%%%%%%%%%%%%%%%%%%%%%%%%%%%%%%%%%%%%%%%%%%%%%%%%%%%%%%%%%%%

%=====================================================================================
\section{Moment sum rule analysis for $c\bar c$ and $b\bar b$ systems} \label{sec:MSR}
%=====================================================================================
For comparison, we may also use the method of the moment sum
rule~\cite{1979-Shifman-p385-447, 1985-Reinders-p1-1} to study the
$2^{--}$ charmonium and bottomonium systems. To suppress the
contribution of the higher states and pick out the lowest lying
resonance, we define the moment by taking derivatives of the
polarization function $\Pi(q^2)$ in Euclidean region $q^2=-Q^2<0$:
\begin{align}
M_n(Q^2_0)=\frac{1}{n!}\bigg(-\frac{d}{dQ^2}\bigg)^n\Pi(Q^2)|_{Q^2=Q_0^2}
=\int_{4m_Q^2}^{\infty}\frac{\rho(s)}{(s+Q^2_0)^{n+1}}ds.
\label{moment}
\end{align}
With the spectral function in Eq.~\ref{Phenrho}, we obtain:
\begin{align}
M_n(Q^2_0)=\frac{f_X^2m_X^6}{(m_X^2+Q_0^2)^{n+1}}\big[1+\delta_n(Q_0^2)\big],
\label{Phemoment}
\end{align}
where $\delta_n(Q_0^2)$ denotes the higher states contributions to
the moment divided by the lowest lying resonance contribution. To
eliminate $f_X$ in Eq.~(\ref{Phemoment}), one can consider the
ratio of the moments:
\begin{eqnarray}
r(n,Q_0^2)\equiv\frac{M_{n}(Q_0^2)}{M_{n+1}(Q_0^2)}=\big(m_X^2+Q_0^2\big)
\frac{1+\delta_{n}(Q_0^2)}{1+\delta_{n+1}(Q_0^2)}. \label{mass2}
\end{eqnarray}
From the ratio $r(n,Q_0^2)$ one can immediately extract the mass
of the lowest lying resonance $m_X$ for
$\delta_{n}(Q_0^2)\cong\delta_{n+1}(Q_0^2)$ at sufficiently high
$n$.

However, taking higher derivative $n$ means moving further away
from the asymptotically free region. This can be compensated by
choosing a larger $Q_0^2$. In Eq.~(\ref{mass2}), it will be
difficult to extract the mass of the lowest lying resonance for
large value of $Q_0^2$ because $\delta_n(Q_0^2)$ converges less
fast in this situation. In fact, one can arrive a region in the
$(n, Q_0^2)$ plane where the lowest lying resonance dominates the
integral in Eq.~(\ref{moment}) and the nonperturbative contribution
is not too large. The moment $M_n(Q^2)$ can be drawn from the
Borel transformed correlation function shown in Sec.~\ref{sec:QSR}
after taking into account the gluon condensate and tri-gluon
condensate.

% %%%%%%%%%%%%%%%%%%%%%%%%%%%%%%%%%%%%%%%%%%%%%%%%%%%%%%%%%%%%%%%%%%%%%%%%%%%%%%
% %\begin{figure}
% \begin{center}
% \begin{tabular}{lr}
% \scalebox{0.60}{\includegraphics{mass1c_n.eps}}&
% \scalebox{0.60}{\includegraphics{mass1b_n.eps}}
% \end{tabular}
% \figcaption{Running masses of the charm and bottom quarks in the moment sum rules for $J/\psi$ and $\Upsilon(1S)$, respectively.} \label{figmoment}
% \end{center}
% %\end{figure}
% %%%%%%%%%%%%%%%%%%%%%%%%%%%%%%%%%%%%%%%%%%%%%%%%%%%%%%%%%%%%%%%%%%%%%%%%%%%%%%

Using the QCD condensates in Eq.~(\ref{parameters}) and the heavy quark masses 
extracted in Fig.~\ref{fig5}, we perform numerical analysis to obtain the $c\bar c$ 
and $b\bar b$ hadron masses as the function of $Q_0^2$ and $n$. In Fig.~\ref{fig71}, 
we show the masses of the
$2^{--}$ charmonium and bottomonium states as the function of $n$.
There is a mass minimum in these curves under the variation with
$n$. By choosing $Q_0^2=50m_c^2, 60m_c^2, 70m_c^2$ for the
charmonium system and $Q_0^2=4m_b^2, 8m_b^2, 12m_b^2$ for the
bottomonium system, the extracted mass converges to $m_{X_{c\bar c}}=(4.13\pm0.20)$ GeV and
$m_{X_{b\bar b}}=(10.27\pm0.39)$ GeV respectively. These values are consistent with the
results of the Borel sum rules within the errors.

%%%%%%%%%%%%%%%%%%%%%%%%%%%%%%%%%%%%%%%%%%%%%%%%%%%%%%%%%%%%%%%%%%%%%%%%%%%%%%
%\begin{figure}
\begin{center}
\begin{tabular}{lr}
\scalebox{0.67}{\includegraphics{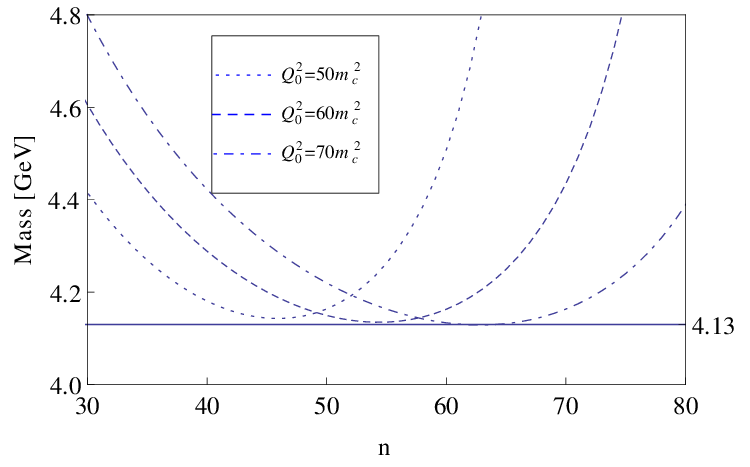}}&
\scalebox{0.67}{\includegraphics{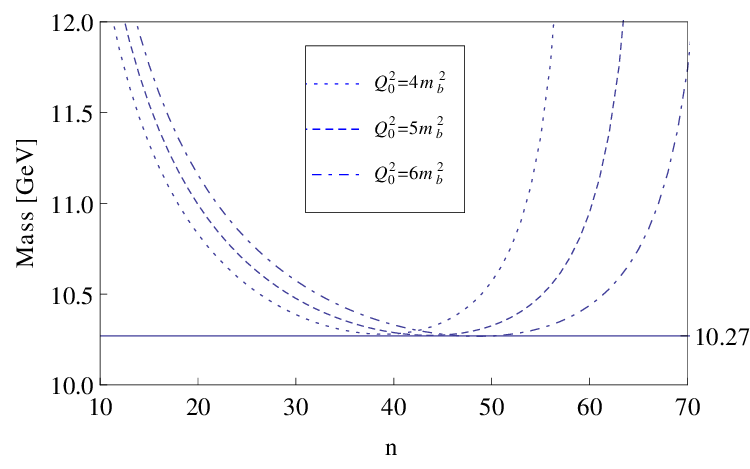}}
\end{tabular}
\figcaption{Variation of $m_X$ with $n$ for the
charmonium and bottomonium states in the framework of
the moment sum rules.} \label{fig71}
\end{center}
%\end{figure}
%%%%%%%%%%%%%%%%%%%%%%%%%%%%%%%%%%%%%%%%%%%%%%%%%%%%%%%%%%%%%%%%%%%%%%%%%%%%%%

%================================================================================
%================================================================================
\section{Summary}\label{sec:summary}
%================================================================================
%================================================================================

We have shown that the interpolating tensor current in Eq.~(\ref{current})
is the correct operator for the $J^{PC}=2^{--}$
meson, the $J_{ij}$ and $J_{0i}$ components of which reduce to the
D-wave and P-wave states respectively in the non-relativistic
limit in terms of the quark model picture as shown in the
appendix. Then we make the operator product expansion (OPE) and
calculate the two-point correlation function. For the heavy quark
$\bar {Q}_1Q_2$ ($Q_1,Q_2=c,b$) systems, the power corrections
include the gluon condensate $\langle g_s^2GG\rangle$ and the
tri-gluon condensate $\langle g_s^3fGGG\rangle$ only. For the
light quarks systems and the light-heavy quarks systems, the four
quark condensate $g_s^2\langle\bar q_1q_1\rangle\langle\bar
q_2q_2\rangle$ (only for the light quarks systems) and the
quark-gluon mixed condensate $\langle \bar qg_s\sigma\cdot
Gq\rangle$ also contribute to the sum rules. While the gluon
condensate $\GGb$ is the dominant nonperturbative correction for
the above systems, these terms also play an important role in the
tensor meson sum rules.

Within the framework of the Borel sum rules, we have studied the
$\bar qq, \bar qs, \bar qc,\bar sc, \bar cc, \bar qb, \bar sb,
\bar cb$ and $\bar bb$ systems. All these systems display stable
QCD sum rules in the working regions of the Borel parameter
$M_B^2$. Up to now, none of these $2^-$ tensor mesons have been
observed except the strange meson $K_2(1770)$ and $K_2(1820)$
~\cite{2010-Nakamura-p75021-75021}. As shown in
Table.~\ref{table1}, the extracted mass of the $\bar qs$ tensor
state is about 1.85 GeV, which is consistent with the mass of the
strange mesons $K_2(1770)$ and $K_2(1820)$
~\cite{2010-Nakamura-p75021-75021}. The lowest D-wave charmonium
state is $\psi(3770)$ with $J^{PC}=1^{--}$. The extracted mass of
the $2^{--}$ D-wave charmonium state is around $3.97$ GeV, which
is slightly higher than $\psi(3770)$ as expected in the quark
model. For the charmonium and bottomonium systems, we also perform
the moment sum rules analysis for comparison. Hopefully the
present investigation will be helpful to the future experimental
search of these tensor states.

%%%%%%%%%%%%%%%%%%%%%%%%%%%%%%%%
\section*{Acknowledgments}
%%%%%%%%%%%%%%%%%%%%%%%%%%%%%%%%
The authors thank Dr. Peng-Zhi Huang and Professor W. Z. Deng for
useful discussions. This project was supported by the National
Natural Science Foundation of China under Grant No. 11261130311.

%\bibliographystyle{h-physrev}
%\bibliography{myreference}

\appendix
%%%%%%%%%%%%%%%%%%%%%%%%%%%%%%%%%%%%%%%%%%%%%%%%%%%%%%%%%%%%%%%%%%%%%%%%%%%%%
\section{THE QUANTUM NUMBERS OF THE INTERPOLATING CURRENT}\label{appendix1}
%%%%%%%%%%%%%%%%%%%%%%%%%%%%%%%%%%%%%%%%%%%%%%%%%%%%%%%%%%%%%%%%%%%%%%%%%%%%%

In order to study the quantum numbers of the tensor current in
Eq.~(\ref{current}), we perform the parity transformation and charge
conjugation transformation to $J_{\mu\nu}$:
\begin{align}
\nonumber\mathbb{P}J_{\mu\nu}\mathbb{P}^{-1}&=-(-1)^{\mu}(-1)^{\nu}J_{\mu\nu}
\\
\mathbb{C}J_{\mu\nu}\mathbb{C}^{-1}&=-J_{\mu\nu}
\end{align}
where $(-1)^{\mu}=1$ for $\mu=0$ and $(-1)^{\mu}=-1$ for
$\mu=1,2,3$. With these relations and the trace condition in
Eq.~(\ref{trace}), the tensor current can couple to the $1^{+-}$
($J_{0i}$ components, i=1,2,3) and $2^{--}$ ($J_{ij}$ components)
states at the same time.

It's interesting to reduce this operator in the non-relativistic
limit and center of mass frame in terms of the quark model
language.
\begin{eqnarray} \nonumber  J_{ij}(x)&=&\bar
Q_1(x)\gamma_{i}\gamma_5\overleftrightarrow{\partial_{j}}Q_2(x)+(i\leftrightarrow
j)
\\
  &\sim&\bar u(p',s)\gamma_{i}\gamma_5ik_{j}v(p,r)+(i\leftrightarrow j)
%   \non&=&(\varphi_s^{\dag},-\varphi_s^{\dag}\frac{\vec{\sigma}\cdot\vec{p'}}{E_{p'}+m})
%   \begin{pmatrix}
%         0&\sigma_{i}\\-\sigma_{i}&0
%   \end{pmatrix}
%   \begin{pmatrix}
%         0&1\\1&0
%   \end{pmatrix}
%   \begin{pmatrix}
%         \frac{\vec{\sigma}\cdot\vec{p}}{E_p+m}\chi_r\\
%         \chi_r
%   \end{pmatrix}
%   (ik_{j})+(i\leftrightarrow j)
%   \non&=&\frac{ik_j}{2m}\varphi_s^{\dag}(\sigma_i\sigma_kp_k+
%   \sigma_k\sigma_ip'_k)\chi_r+(i\leftrightarrow j)
%   \non&=&\frac{ik_j}{2m}\varphi_s^{\dag}[(\delta_{ik}+i\epsilon_{ikl}\sigma_l)p_k+
%   (\delta_{ki}+i\epsilon_{kil}\sigma_l)p'_k]\chi_r+(i\leftrightarrow j)
%   \non&=&\frac{i}{2m}\varphi_s^{\dag}[k_jq_i+i\epsilon_{ikl}\sigma_lk_kk_j]\chi_r+(i\leftrightarrow j)
=-\frac{1}{2m}\varphi_s^{\dag}\epsilon_{ikl}\sigma_lk_kk_j\chi_r+(i\leftrightarrow j)
\non
 J_{0i}(x)&=&\bar Q_1(x)(\gamma_{0}\gamma_5\overleftrightarrow{\partial_{i}}+
  \gamma_{i}\gamma_5\overleftrightarrow{\partial_{0}})Q_2(x)
\\
  &\sim&\bar u(p',s)\gamma_{0}\gamma_5ik_{i}v(p,r)
%   \non&=&(\varphi_s^{\dag},\varphi_s^{\dag}\frac{\vec{\sigma}\cdot\vec{p'}}{E_{p'}+m})
%   \begin{pmatrix}
%         0&1\\1&0
%   \end{pmatrix}
%   \begin{pmatrix}
%         \frac{\vec{\sigma}\cdot\vec{p}}{E_p+m}\chi_r\\
%         \chi_r
%   \end{pmatrix}
%   (ik_{i})
%   \non&=&ik_{i}\varphi_s^{\dag}[\frac{(\vec{\sigma}\cdot\vec{p})(\vec{\sigma}\cdot\vec{p'})}{4m^2}+1]\chi_r
=ik_{i}\varphi_s^{\dag}\chi_r
\\
J_{00}(x)&\sim& ik_{0}\varphi_s^{\dag}\chi_r=0
\end{eqnarray}
in which $k=p-p'$ and  ${\vec q}={\vec p}+{\vec p}'=0$ in the
center of mass system. We have used the non-relativistic limit:
$E_{p'}=E_p=m, k_0=0$. It is obvious that $J_{ij}$ reduces to the
D-wave and $J_{0i}$ reduces to P-wave in the non-relativistic
limit. Therefore the quantum numbers of the current should be
$2^{--}$ for the $J_{ij}$ component and $1^{+-}$ for the $J_{0i}$
component.

%%%%%%%%%%%%%%%%%%%%%%%%%%%%%%%%%%%%%%%%%%%%%%%%%%%%%%%%%%%%%%%%%%
\section{THE MOMENTUM SPACE PROPAGATOR}\label{appendix2}
%%%%%%%%%%%%%%%%%%%%%%%%%%%%%%%%%%%%%%%%%%%%%%%%%%%%%%%%%%%%%%%%%%

To calculate the higher dimensional gluonic operators, we consider
the gluon field as an external one with the fixed-point gauge
condition~\cite{1981-Dubovikov-p109-132, 1980-Cronstrom-p267-269,
1980-Shifman-p13-13}:
\begin{align}
(x-x_0)^{\mu}A_{\mu}^a(x)=0, \label{gauge}
\end{align}
where $x_0$ is an arbitrary point in space which can be chosen to
be the origin. Then the four potential $A_{\mu}^a$ can be
expressed in terms of the field strength tensor
$G_{\mu\nu}$($G_{\mu\nu}=\frac{\lambda^a}{2}G_{\mu\nu}^a$):
\begin{align}
\nonumber  A_{\mu}(x)&=\int_0^1tdtG_{\nu\mu}(tx)x^{\nu} \non
&=\frac{1}{2}x^{\nu}G_{\nu\mu}(0)+\frac{1}{3}x^{\alpha}x^{\nu}D_{\alpha}G_{\nu\mu}(0)
+\frac{1}{8}x^{\alpha}x^{\beta}x^{\nu}D_{\alpha}D_{\beta}G_{\nu\mu}(0)+...,
\end{align}
%%%%%%%%%%%%%%%%%%%%%%%%%%%%%%%%%%%%%%%%%%%%%%%%%%%%%%%%%%%%%%%%%%%%%%%%%%%%%%
%\begin{figure}
\begin{center}
\begin{tabular}{l}
\scalebox{0.5}{\includegraphics{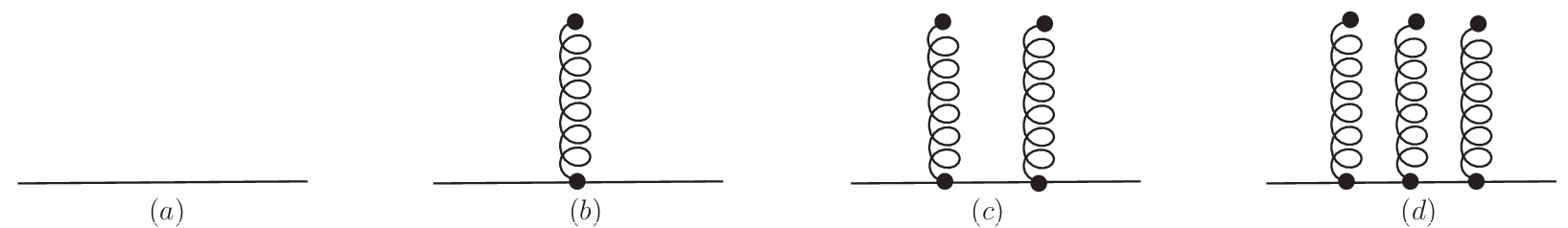}}
\end{tabular}
\figcaption{Graphic representations of the massive quark
propagators with gluon legs attached to the quark line.}
\label{figpropa}
\end{center}
%\end{figure}
%%%%%%%%%%%%%%%%%%%%%%%%%%%%%%%%%%%%%%%%%%%%%%%%%%%%%%%%%%%%%%%%%%%%%%%%%%%%%%
Denote the massive quark propagator between the position $x$ and
$y$ in the coordinate space as $iS_0(x, y)$. The massive quark
propagator in the momentum space can be obtained as:
\begin{align}
  iS(p) &=iS_0(p)+iS_g(p)+iS_{gg}(p)+iS_{ggg}(p),
\end{align}
where $iS_0(p)$ is the free quark propagator as shown in
Fig.~\ref{figpropa}(a):
\begin{align}
  iS_0(p)=\frac{i}{\hat{p}-m},
\end{align}
$iS_g(p)$ is the quark propagator with one gluon leg attached as
shown in Fig~\ref{figpropa}(b):
\begin{eqnarray}
iS_g(p)=\frac{i}{4}\frac{\lambda^n}{2}g_sG^n_{\mu\nu}\frac{\sigma^{\mu\nu}(\hat{p}+m)+(\hat{p}+m)\sigma^{\mu\nu}}
{(p^2-m^2)^2}+\frac{i}{2}\frac{\lambda^n}{2}g_sG^n_{\mu\nu}\bigg[\frac{2y^{\mu}p^{\nu}(\hat{p}+m)}
{(p^2-m^2)^2}-\frac{y^{\mu}\gamma^{\nu}}{p^2-m^2}\bigg]
\end{eqnarray}
$iS_{gg}(p)$ is the quark propagator with two gluon legs attached
as shown in Fig~\ref{figpropa}(c):
\begin{eqnarray}
\nonumber
iS_{gg}(p)&=&-\frac{i}{4}\frac{\lambda^a}{2}\frac{\lambda^b}{2}g^2_sG^a_{\mu\rho}G^b_{\nu\sigma}
\frac{\hat{p}+m}{(p^2-m^2)^5}\big(f^{\mu\rho\nu\sigma}+f^{\mu\nu\rho\sigma}+f^{\mu\nu\sigma\rho}\big)
\\
&&-\frac{1}{4}\frac{\lambda^a}{2}\frac{\lambda^b}{2}g^2_sG^a_{\mu\rho}G^b_{\nu\sigma}
\frac{\hat{p}+m}{(p^2-m^2)^4}\Big[y^{\sigma}(f^{\mu\rho\nu}+f^{\mu\nu\rho})+y^{\rho}f^{\mu\nu\sigma}
-iy^{\rho}y^{\sigma}f^{\mu\nu}(p^2-m^2)\Big]
\end{eqnarray}
$iS_{ggg}(p)$ is the quark propagator with three gluon legs
attached as shown in Fig~\ref{figpropa}(d):
\begin{eqnarray}
\nonumber
iS_{ggg}(p)&=&\frac{i}{8}\frac{\lambda^a}{2}\frac{\lambda^b}{2}\frac{\lambda^c}{2}g_s^3G^a_{\mu\alpha}G^b_{\nu\beta}G^c_{\rho\gamma}
\frac{\hat{p}+m}{(p^2-m^2)^7}
\big(f^{\mu\alpha\nu\beta\rho\gamma}+f^{\mu\alpha\nu\rho\beta\gamma}+f^{\mu\alpha\nu\rho\gamma\beta}
+f^{\mu\nu\alpha\beta\rho\gamma}+f^{\mu\nu\beta\alpha\rho\gamma}+f^{\mu\nu\beta\rho\alpha\gamma}
\non&&+f^{\mu\nu\beta\rho\gamma\alpha}+f^{\mu\nu\alpha\rho\beta\gamma}+f^{\mu\nu\alpha\rho\gamma\beta}
+f^{\mu\nu\rho\alpha\beta\gamma}+f^{\mu\nu\rho\beta\alpha\gamma}
+f^{\mu\nu\rho\beta\gamma\alpha}+f^{\mu\nu\rho\alpha\gamma\beta}+f^{\mu\nu\rho\gamma\alpha\beta}+f^{\mu\nu\rho\gamma\beta\alpha}\big)
\non&&
-\frac{i}{8}\frac{\lambda^a}{2}\frac{\lambda^b}{2}\frac{\lambda^c}{2}g_s^3G^a_{\mu\alpha}G^b_{\nu\beta}G^c_{\rho\gamma}
\frac{\hat{p}+m}{(p^2-m^2)^6}\bigg\{\Big[iy^{\alpha}(f^{\mu\nu\beta\rho\gamma}+f^{\mu\nu\rho\beta\gamma}+f^{\mu\nu\rho\gamma\beta})
+iy^{\beta}(f^{\mu\alpha\nu\rho\gamma}+f^{\mu\nu\alpha\rho\gamma}
\non&&+f^{\mu\nu\rho\gamma\alpha}+f^{\mu\nu\rho\alpha\gamma})
+iy^{\gamma}(f^{\mu\alpha\nu\beta\rho}+f^{\mu\alpha\nu\rho\beta}+f^{\mu\nu\alpha\beta\rho}+f^{\mu\nu\beta\alpha\rho}
+f^{\mu\nu\beta\rho\alpha}+f^{\mu\nu\alpha\rho\beta}+f^{\mu\nu\rho\alpha\beta}+f^{\mu\nu\rho\beta\alpha})\Big]
\non&&+(p^2-m^2)\Big[y^{\alpha}y^{\beta}f^{\mu\nu\rho\gamma}+y^{\alpha}y^{\gamma}(f^{\mu\nu\beta\rho}+f^{\mu\nu\rho\beta})+
y^{\beta}y^{\gamma}(f^{\mu\alpha\nu\rho}+f^{\mu\nu\alpha\rho}+f^{\mu\nu\rho\alpha})\Big]
\\&&
-iy^{\alpha}y^{\beta}y^{\gamma}f^{\mu\nu\rho}(p^2-m^2)^2\bigg\}
\end{eqnarray}
where
$f^{\mu\nu...\alpha\beta}=\gamma^{\mu}(p^2-m^2)\gamma^{\nu}(p^2-m^2)...\gamma^{\alpha}(p^2-m^2)\gamma^{\beta}(p^2-m^2)$.

%===================================================================================================================================
%===================================================================================================================================

\end{document}